\documentclass[10pt,letterpaper]{article}
\usepackage{opex3}
\usepackage{color}
\usepackage{amsmath}
\usepackage{bbold}
\begin{document}

\title{Quantum correlation of light scattered by disordered media}

\author{Ilya Starshynov$^{*}$, Jacopo Bertolotti, Janet Anders} 

\address{ University of Exeter, Stocker Road, Exeter EX4 4QL, United Kingdom}

\email{$^*$is283@exeter.ac.uk} 

\begin{abstract}
We study theoretically how multiple scattering of light in a disordered medium 
can spontaneously generate quantum correlations. In particular we focus on the 
case where the input state is 
Gaussian and characterize the correlations between two arbitrary output modes. 
As there is not a single all-inclusive measure of correlation, we characterise 
the output correlations with three measures:
 intensity fluctuations, entanglement, and 
quantum discord. We found that, while a single mode coherent state input can not
produce quantum correlations, any other Gaussian input will produce them in 
one form or another. This includes input states that are usually regarded as 
more classical 
than coherent ones, such as thermal states, which will produce a non zero
quantum discord.
\end{abstract}


\section{Introduction}

Light propagation in a scattering medium can often be described as a diffusive 
process \cite{Sheng}, where any memory of the initial state is lost almost 
immediately, and transport is represented by the incoherent sum over many 
Brownian random walks. However it was early realized that temporal coherence 
survives multiple elastic scattering, and thus interference is still possible 
even after passing through a diffusive medium \cite{Goodman}. As a consequence 
of interference the diffusive picture has to be modified, and the light 
scattered into different channels develop correlations \cite{Akkermans}. 
Classical correlations between the intensity scattered in one direction and the 
intensity scattered into another direction were studied since the '60s 
\cite{Stone, Feng} and found many applications, especially in imaging, e.g. in 
dynamic light scattering \cite{DLS}, speckle contrast imaging \cite{Dunn}, 
stellar speckle interferometry \cite{Dainty}, wavefront shaping \cite{Mosk}, 
and speckle scanning microscopy \cite{Bertolotti, Gigan}. The study of quantum 
correlations of multiply scattered light started much later, probably due to 
the widespread idea that quantum features are frail and thus unlikely to play a 
significant role in the presence of strong disorder. 
Nevertheless it was recently shown that not only certain quantum features can 
survive multiple scattering \cite{Lodahlnoise}, but that quantum correlations 
can even be spontaneously created in the scattering process 
\cite{Lodahlentanglement}. The quantum properties of scattered light proved to 
be so robust that control over the transport of single photons in a disordered 
medium via wavefront shaping was demonstrated \cite{Pinkse, Hugo}.

In this paper we consider the case of a generic Gaussian input state and study 
theoretically the necessary conditions for the output state to present quantum 
correlations. In particular we show that, while entanglement requires specific 
conditions to be produced, quantum discord is always present in the output 
state when a thermal input state is used but, surprisingly, not when the input 
mode is in a coherent state.

\section{Quantum correlations between two modes}

Classical correlations of scattered light are commonly described by a
correlation function $\mathcal{C}_{l,m}$ that measures the correlations between
intensity fluctuations of two different modes $l$ and $m$ 
\cite{Akkermans, Stone, Feng},
\begin{equation}
\label{eq:classicalcorr}
	 \mathcal{C}_{l,m} =
    \frac{\left\langle I_{l} I_{m} \right\rangle}
    {\left\langle I_{l} \right\rangle \left\langle I_{m} \right\rangle},
\end{equation}
where $I=\left|E\right|^2$ is the light intensity and $\left\langle \cdot 
\right\rangle$ represents either a time or an ensemble average (if the system 
is ergodic the two are  equivalent). $\mathcal{C}$ can be used beyond the 
classical case to study certain classes of quantum correlations by substituting 
$I$ with the mode's number operator $\hat{n}$ \cite{Glauber, Lodahl2005}, and 
since classical light can never lead to $\mathcal{C}<1$, a correlation value 
below 1 is considered a clear signature of quantumness \cite{loudon}.

In order to study the effect of multiple scattering on a Gaussian input state 
we employ the formalism of density operators and covariance matrices 
\cite{Laurat2005}. The covariance matrix of two output modes is defined as
\begin{equation} \label{eq:CovarianceMatrix}
\sigma_{\mu,\nu} = \frac{1}{2} \langle\hat{R}_{\mu} \hat{R}_{\nu} +
\hat{R}_{\nu} \hat{R}_{\mu} \rangle - \langle \hat{R}_{\mu} \rangle \langle
\hat{R}_{\nu} \rangle , \quad \mu, \nu
= 1, ..., 2N ,
\end{equation}
where  $\hat{R} = \{\hat{x}_1,\hat{p}_1 \dots \hat{x}_N,\hat{p}_N \}$ is
the vector of quadrature operators, that satisfies the commutation relation $ [ 
 \hat{R}_{\mu},  \hat{R}_{\nu}] = i  \, \left( \Omega^{\oplus
N} \right)_{\mu,\nu} $, which imposes constrains on $\sigma$:
\begin{equation}
\label{eq:commutation}
  \sigma + i \Omega^{\oplus
N}  \ge 0 \quad \mbox{with} \quad
    \Omega =
    \begin{pmatrix}
   	 0 & 1\\
   	 -1 & 0
    \end{pmatrix},
\end{equation}
where $\oplus$ is a direct sum. Importantly, for Gaussian states the covariance 
matrix captures all the correlation properties between the modes. As will be 
discussed below, 
different measures of quantum correlations, such as entanglement and discord, 
can be formulated as conditions on the elements of the covariance matrix.

\begin{figure}[t]
    \centering
    \includegraphics[scale=1]{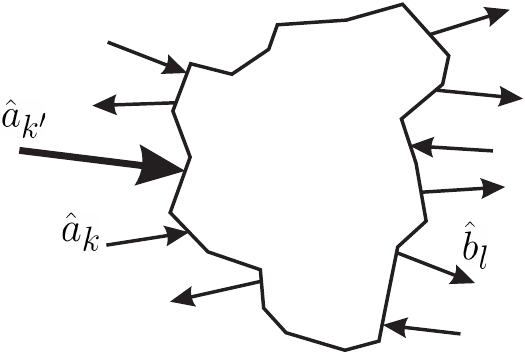}
    \caption{ \label{fig:scattering}The disordered medium is scattering $N$
input modes $\hat{a}_{k}$ into $N$ output modes $\hat{b}_{l}$. The scattering
process is described by a scattering matrix $S$. The only non-empty input mode
is $\hat{a}_{k^{\prime}}$, the others are assumed to be in a vacuum state.}
\end{figure}
When monochromatic light propagates through a disordered medium it is multiply
scattered, and its wavefront becomes completely irregular. As a result the 
output
modes take the form of a speckle pattern \cite{Goodman}. Here we consider a 
model of linear scattering, such that light propagation can be described
by the scattering matrix $S$ \cite{Beenakker1997}, which couples the fields of
the $N$ output modes, $E^{out}_l$,  with the $N$ input modes, $E^{in}_k$ 
\cite{Lodahl2005}. The electromagnetic field of a mode with polarization 
$\mathbf{e}_k$, wave vector $\mathbf{q}_k$ and frequency $\omega_k$ can be 
expressed
as:
\begin{equation}
E^{in}_k(\mathbf{r})= \mathbf{e}_k \sqrt{\frac{2 \hbar \omega_k}{2 \epsilon_0 
V}} 
\left[ \hat{a}_k e^{-i\mathbf{q}_k \mathbf{r} } + \hat{a}^\dagger_k 
e^{i\mathbf{q}_k\mathbf{r}}\right],
\end{equation}
where  $\hat{a_k}$ and $\hat{a}_k^\dagger$ are the creation and annihilation 
operators of the input mode, $\epsilon_0$ is the vacuum permittivity and $V$ is 
the 
mode volume~\cite{loudon}.
The scattering matrix $S$ links the input fields to the output ones as 
$E^{out}_l=\sum_k^N S_{l,\, k}\, E^{in}_k $. Likewise for the ladder operators 
we have:
\begin{equation} \label{eq:in-out}
\hat{b}_l=\sum_{k}^{N}
S_{l,k} \, \hat{a}_k, \quad k, l, m = 1..N.
\end{equation}
Here $\hat{a}$ and $\hat{b}$ are the annihilation operators related to,
respectively, the input and the output modes as shown in
Fig.~\ref{fig:scattering}. The elements of the $S$ matrix are the complex 
transmission coefficients from the $k$-th input mode to the $l$-th output mode. 

In this paper we focus on the correlations between two of
the $N$ scattered modes, whose covariance matrix $\sigma$ has the structure:
\begin{equation} \label{ABCCT}
\sigma = \begin{pmatrix}
A & \Gamma\\
\Gamma^T &B
\end{pmatrix},
\end{equation}
where $A$, $B$, and $\Gamma$ are $2\times2$ matrices.
The matrices $A$ and $B$ refer to individual
properties of the two modes, and the matrix $\Gamma$ describes the correlations
between them. The determinants of these matrices, together with the determinant
of the whole covariance matrix, are invariant under local transformations
(i.e. those acting only on one of the modes). Because of
this local invariance the determinants characterize entanglement and other
non-local correlation properties of the state 
\cite{Adesso2010,Peres1996,Horodecki1996,Simon1994}.
 Such two-mode covariance matrix can be written in the form
\begin{equation} \label{eq:simon}
\sigma = \begin{pmatrix}
 \alpha& 0 & \gamma_x & 0\\
 0 & \alpha & 0 & \gamma_p\\
 \gamma_x& 0 & \beta & 0\\
 0& \gamma_p & 0 & \beta
\end{pmatrix}
\end{equation}
keeping the invariants of the state: $\det(A) =\alpha^2, \, \det(B) = \beta^2, 
\, \det(\Gamma) = \gamma_x \gamma_p$ and $\det(\sigma)$ the same 
\cite{Simon1999}.

The intensity correlation $\mathcal{C}$ contains fourth order moments of the
field distribution. For Gaussian states these can always be written as a 
function of the
second order moments, contained in the covariance matrix $\sigma$
\cite{Laurat2005}. This leads to the expression for the correlation function:
\begin{equation} \label{correl_C}
\mathcal{C} =1+ \frac{ (2\gamma_x^2+2\gamma_p^2)}{(2\alpha-1)(2\beta-1)} .
\end{equation}
From Eq.~\eqref{correl_C} we can see that Gaussian states always have
$\mathcal{C} \ge 1$, as $\alpha,\beta \ge 1/2$ as a consequence of the
commutation relations in Eq.~\eqref{eq:commutation}. 

While $\mathcal{C}$ is useful to describe classical correlations and certain 
features of quantum light, like photon anti-bunching \cite{Glauber}, it does 
not capture other quantum correlations, and thus other measures have 
been introduced, notably entanglement and discord. The characterization of 
entanglement for general mixed states of multiple modes is a challenging task. 
Necessary and sufficient conditions for entanglement have been identified for 
bipartite discrete quantum systems of $2\times2$ and $2\times3$ 
dimensions \cite{Peres1996,Horodecki1996} and for continuous two-mode Gaussian 
states 
\cite{Simon1999,Duan1999}. In the latter case, 
these mathematical conditions can be understood as inverting time in one of the 
modes of the system and checking if the resulting quantum state is still a 
valid 
physical state, i.e. 
hermitian and positive~\cite{Peres1996,Horodecki1996,Simon1999,Duan1999}. These 
conditions on the quantum state can be recast as conditions
on the matrix $\sigma^\prime$, obtained from the covariance 
matrix $\sigma$ Eq.~\eqref{eq:simon}, by changing the sign to the 
momentum of one of the two modes, 
i.e. $\sigma^{\prime} = \sigma (\gamma_p \to - \gamma_p)$. If $\sigma^{\prime}$ 
obeys the commutation relation stated in Eq.~\eqref{eq:commutation} then the 
two modes are separable, otherwise they are entangled. This separability 
condition for the two modes can  be rewritten as \cite{Adesso2010},
\begin{equation}\label{eq:determ}
	\eta^{\pm}:=\left[\frac{1}{2}\left(\Delta^{\prime} \pm
		\sqrt{\Delta^{\prime 
2}-4\det(\sigma^{\prime})}\right)\right]^{1/2} \ge 1/2,
\end{equation}
where $\Delta^{\prime} = \det(A)+\det(B)+2\det(\Gamma^{\prime})$ refers to the 
sub-matrices of the covariance matrix $\sigma^{\prime}$ and 
$\det(\Gamma^{\prime}) = - \gamma_p \gamma_x$ (see Eq.~\eqref{ABCCT} and 
Eq.~\eqref{eq:simon}).  $\eta^{\pm}$ are the symplectic eigenvalues of the 
matrix $\sigma^{\prime}$ \cite{Laurat2005}.

\begin{figure}[b!]
	\centering
	\includegraphics[scale=1]{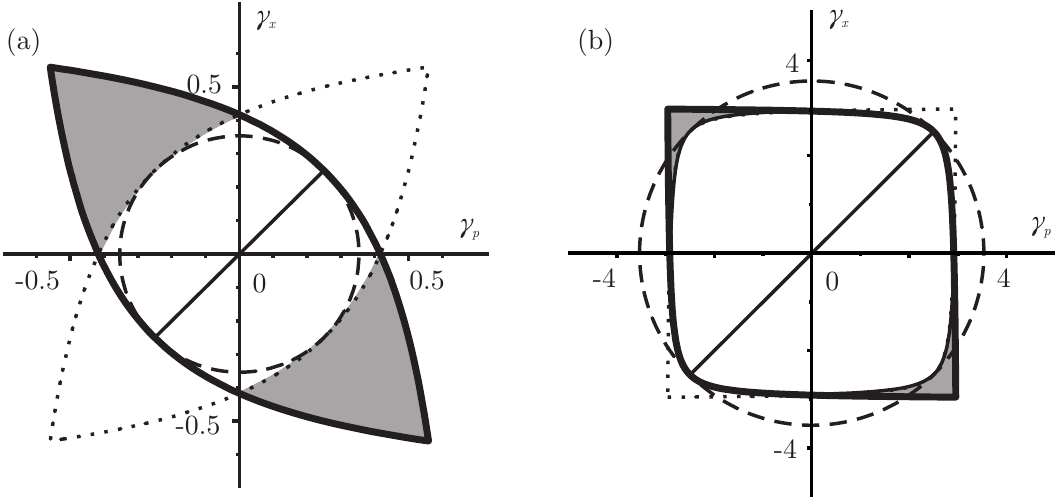}
	\caption{ \label{fig:regions} Map of all possible two-mode Gaussian 
states and their correlations, as a function of the 
off-diagonal elements 
$\gamma_x$ and $\gamma_p$ of the covariance matrix in Eq.~\eqref{eq:simon}, 
with diagonal values $\alpha$ and $\beta$: (a) $\alpha=\beta=0.75$,  (b) 
$\alpha 
= \beta=5$. All 
physically allowed states lie within the region enclosed by the thick solid  
line. The region of separable states can be found as the intersection of the 
allowed state region and its mirror image with respect of $\gamma_p$ (dotted 
line)~\cite{Simon1999}. The remaining states (in the gray area) are not 
separable, i.e. they are entangled.
For a thermal input state the output states, Eq.~\eqref{eq:term}, lie on the 
$\gamma_x = \gamma_p$ line, and thus are not entangled. The dashed circle 
corresponds to the condition 
$\mathcal{C}=2$. Inside this circle $\mathcal{C}$ 
is less
 than 2, with $\mathcal{C}=1$ at the origin. 
 Increasing the number of photons in the input state, i.e. increasing $\alpha$ 
and $\beta$, the fraction of entangled output states decreases and some 
entangled states cross into the $\mathcal{C}=2$ circle,
as shown in panel (b).} 
\end{figure}

When the parameters $\alpha$ and $\beta$ (i.e. the number of
photons in each of the modes) are fixed, 
Eq.~\eqref{eq:commutation}
and Eq.~\eqref{eq:determ} define closed regions in the $\gamma_x$ and
$\gamma_p$
parameter space (see Fig.~\ref{fig:regions}). The condition in 
Eq.~\eqref{eq:commutation} defines
an area enclosed by the thick solid line in 
Fig.~\ref{fig:regions}, which corresponds to all valid states. The entangled 
states lie within this areas but outside the region defined by the condition in
Eq.~\eqref{eq:determ} (gray areas in Fig.~\ref{fig:regions}), which can be 
graphically obtained
as the first region rotated by $90^{\circ}$ (dotted lines in
Fig.~\ref{fig:regions}).

For a long time entanglement was considered to be the quintessential 
non-classical ingredient for quantum computation and communication.
However, it was shown that even using separable (i.e. non-entangled) states it 
is possible to perform certain computational tasks exponentially faster than 
any 
known classical algorithm \cite{Datta2005}. 
This means that just the fact of the system being quantum may 
lead to applications impossible for classical systems, irrespectively of 
separability. This fact has been 
formalized by introducing quantum discord \cite{Ollivier2001}, a measure which 
broadens the concept of non-classical correlations to states without 
entanglement.

For a pure bipartite entangled state the measurement of one of the parties 
completely determines the result of the measurement on the second, while for 
classical states measurement on one party does not effect the other. On the 
other 
hand there are separable states where the measurement of one party influences 
the other in a probabilistic sense \cite{Rahimi2013,Hosseini2014}, making them 
non-classical, but not entangled either. Quantum discord quantifies this 
influence and is defined as the discrepancy of two measures of 
mutual information for a joint quantum state $\rho_{AB}$ of systems $A$ and $B$:
\begin{equation}\label{mutInfo}
I^q(\rho_{AB})=S(\rho_A)+S(\rho_B)-S(\rho_{AB})\quad \text{and} \quad 
J^q(\rho_{AB}) = S(\rho_A)-\sum_j Tr[\rho_{AB}  \Pi_j ] \, \, S(\rho_{A| \Pi_j} 
).
\end{equation}
Here  $\rho_A = Tr_B[\rho_{AB}]$ and  $\rho_B = Tr_A[\rho_{AB}]$ are the 
reduced 
states of $\rho_{AB}$ and $S$  is the von Neumann entropy: $S(\rho)=-Tr[\rho 
\ln \rho ]$. 
The second equation depends on the measurement choice $\{ \Pi_j \}$, and 
$\rho_{A| \Pi_j} =Tr_B[\rho_{AB} \Pi_j ]/Tr[\rho_{AB}  \Pi_j ]$ 
is the state of the system A conditioned on a certain outcome of this 
measurement performed on the subsystem B.

Quantum discord $D(\rho_{AB})$ is quantified by the difference between the two 
expressions in Eq.~\eqref{mutInfo} minimized over all possible sets of 
measurement 
operators,
\begin{equation} \label{discord}
D(\rho_{A|B})=\inf_{\Pi_j}\left( I^q(\rho_{AB}) - J^q(\rho_{AB}) 
\right)=S(\rho_B)-S(\rho_{AB})+\inf_{ \Pi_j}\sum_j Tr[\rho_{AB} 
 \Pi_j] \, S(\rho_{A| \Pi_j}).
\end{equation}
When the input state is Gaussian and its covariance matrix is 
$\sigma$, the expression for the entropy of this state reduces to: 
$S(\sigma)=\sum_i \kappa(\eta_i)$, where $\eta_i$ are the symplectic 
eigenvalues 
of $\sigma$ and $\kappa(z)=  (z+1/2)\ln(z+1/2)+(z-1/2)\ln(z-1/2)$ 
\cite{Holevo2001}.
Here we limit ourselves to Gaussian measurements, i.e. those that preserve the 
Gaussian nature of the state they are applied to. Under this assumptions the 
minimization of Eq.~\eqref{discord}, written as a function of the invariants of 
the covariance matrix $\sigma$ from Eq.~\eqref{eq:simon}, gives:
\begin{equation} \label{eq:discord}
D_G(\sigma)=\kappa(\sqrt{\det(B)})-\kappa(\eta^-)-\kappa(\eta^+)+\kappa\left( 
\frac{\sqrt{\det(A)}+2\sqrt{\det(A)\det(B)}+2\det(\Gamma)}{1+2\sqrt{\det(B)}}
\right),
\end{equation}
which is known as Gaussian discord \cite{Giorda2010}. The Gaussian discord 
$D_G$ vanishes if and only if $\Gamma = 
0$~\cite{Adesso2010,Rahimi2013,Giorda2010}.

\section{Correlations between two scattered modes}

To characterize the correlation of two modes of the scattered light we 
explicitly calculate the elements of the corresponding covariance matrix. We 
will consider the experimentally common situation where the input light is in a 
single mode $k^{\prime}$, and all the other input modes are in a vacuum state, 
as depicted in Fig.~\ref{fig:scattering}. 
Replacing the quadrature operators in the vector $\hat{R}$ with the ladder 
operators:
$\hat{x} = (\hat{b} + \hat{b}^{\dagger})/\sqrt{2}$ and $\hat{p} = (\hat{b} 
-\hat{b}^{\dagger})/i\sqrt{2} $, and substituting Eq.~\eqref{eq:in-out} in, the
general expression for the elements of the covariance matrix of two output 
modes is
\begin{equation}
\begin{aligned}
\sigma_{2l-1,2m-1} &=
\frac{\delta_{l,m}}{2}
+W_{l,m}\, \Delta \hat{n}_{k^{\prime}}+Y_{l,m}\, \Delta \hat{a}_{k^{\prime}} 
\hat{a}_{k^{\prime}}
+Y^{*}_{l,m} \, \Delta 
\hat{a}_{k^{\prime}}^{\dagger}\hat{a}_{k^{\prime}}^{\dagger}, \\
\sigma_{2l,2m} &=
\frac{\delta_{l,m}}{2}
+W_{l,m} \, \Delta \hat{n}_{k^{\prime}}-Y_{l,m} \, \Delta \hat{a}_{k^{\prime}}  
\hat{a}_{k^{\prime}}  
-Y^{*}_{l,m} \, \Delta 
\hat{a}_{k^{\prime}}^{\dagger}\hat{a}_{k^{\prime}}^{\dagger}, \\
\sigma_{2l-1,2m}=\sigma_{2l,2m-1} &=
\frac{1}{2i}\big[
Z_{l,m} \, \Delta \hat{n}_{k^{\prime}}
+Y_{l,m} \, \Delta \hat{a}_{k^{\prime}}  \hat{a}_{k^{\prime}}
-Y^{*}_{l,m} \, \Delta 
\hat{a}_{k^{\prime}}^{\dagger}\hat{a}_{k^{\prime}}^{\dagger} \big] ,
\end{aligned}
\label{eq:CM_gen}
\end{equation}
where
\begin{equation*}
\begin{aligned}
\Delta \hat{n}_{k^{\prime}} &=\langle \hat{a}^{\dagger}_{k^{\prime}}  
\hat{a}_{k^{\prime}} \rangle - \langle  \hat{a}^{\dagger}_{k^{\prime}} \rangle
\langle  \hat{a}_{k^{\prime}} \rangle , \\
\Delta \hat{a}_{k^{\prime}} \hat{a}_{k^{\prime}} &= \langle  
\hat{a}_{k^{\prime}} 
\hat{a}_{k^{\prime}} \rangle - \langle  \hat{a}_{k^{\prime}} \rangle \langle  
\hat{a}_{k^{\prime}} \rangle \\
W_{l,m}& = \left( S^{*}_{l,k^{\prime}}S_{m,k^{\prime}} +
S^{*}_{m,k^{\prime}}S_{l,k^{\prime}} \right), \\
Z_{l,m} &=  \left( S^{*}_{l,k^{\prime}}S_{m,k^{\prime}} -
S^{*}_{m,k^{\prime}}S_{l,k^{\prime}} \right), \\
Y_{l,m} &= S_{l,k^{\prime}}S_{m,k^{\prime}},
\end{aligned}
\end{equation*}
This expression allows us to analyse
the correlation properties between two
output modes, e.g. entanglement and quantum discord, for 
different possible inputs.

\paragraph{Coherent state:}
If the input mode is in a coherent state  Eq.~\eqref{correl_C} gives the
value of  $\mathcal{C}_{l,m} =1$ as expected, which means 
that there is no correlation between the intensity fluctuations of the two 
output modes according to this measure~\cite{MandelBook}.
Since all expectation values of the operators $\Delta \hat{n}_{k^{\prime}}$ and 
$\Delta \hat{a}_{k^{\prime}} \hat{a}_{k^{\prime}}$ are 0, the covariance matrix 
of the output modes will be $\sigma^{coh} = \frac{1}{2}  
\mathbb{1}$. Substituting  the elements 
of $\sigma^{coh}$ into Eq.~\eqref{eq:determ}, the allowed 
region shrinks to a point $\gamma_x=~\gamma_p=0$. This means that for any
coherent state as an input, 
any two output modes will simply be a product of two coherent states 
and no quantum correlations are present. 

\paragraph{Squeezed state:}
If the input mode is in a squeezed state with a squeezing parameter $r$ and 
phase $\theta$, the expectation values for the operators in 
Eq.~\eqref{eq:CM_gen} are:   $\Delta \hat{a}_{k^{\prime}} \hat{a}_{k^{\prime}} 
=-e^{i\theta}\sinh(r) \cosh(r)$,  and $\Delta \hat{n}_{k^{\prime}} 
=\sinh^{2}(r)$ \cite{Teich1989}. If we set $\theta=0$, for 
which we expect maximal entanglement \cite{Laurat2005}, we can express  the 
entanglement criterion 
as $\sinh(r)^{2}|S_{l,k^{\prime}}|^{2}|S_{m,k^{\prime}}|^{2}>0$, which is 
always 
true if $r$ and both scattering matrix elements are non-zero. 
This means that we will get entanglement for any non-zero 
degree of squeezing $r$ \cite{Paris1999}. It is remarkable that the degree of 
entanglement does not depend on the phases of transmission coefficients of the 
scattering matrix, but only on their moduli.

Although quantum features of intensity fluctuation correlations are often 
linked to $\mathcal{C}<1$, implying a reduction of 
coincidences in simultaneous detections of photons in the two modes, for 
squeezed states entanglement leads to positive correlation, which can lead to 
high $\mathcal{C}$.
The maximal possible value of $\mathcal{C}$ which could be achieved with 
squeezed entangled states (the top left or bottom right points of the region of 
allowed states in Fig.~\ref{fig:regions}) is:
\begin{equation}
\mathcal{C}^{sq}_{l,m} = 
2+\frac{|S_{l,k^{\prime}}|^2+|S_{m,k^{\prime}}|^2}{2\bar{n}|S_{l,k^{\prime}}||S_
{m,k^{\prime}}|},
\end{equation}
where $\bar{n}$ is the average number of photons in the input mode. When 
$\bar{n} \to \infty$, $\mathcal{C}^{sq}$ approaches 2, which corresponds to the 
value expected for thermal states. The presence of entanglement allows to reach 
values of $\mathcal{C}$  inaccessible for thermal states, and this is exploited 
in quantum imaging where it can allow faster recovery of information, 
especially 
in the low photon number regime \cite{Gatti2004}. However, this needs to be 
treated with some care. In fact, as it can be seen from 
Fig.~\ref{fig:regions}b, 
with the increase of $\alpha$ and $\beta$ the circle $\mathcal{C}=2$ and 
the boundary of the gray area (corresponding to the entangled states) cross, 
and 
therefore it is possible to find non-entangled states with higher intensity 
correlations than some entangled states.

\paragraph{Thermal state:} 
For a thermal state $\rho_{th}$ we have: $Tr(\hat{a}_{k^{\prime}} \rho_{th}) = 
Tr(\hat{a}_{k^{\prime}}^{\dagger} \rho_{th}) = 
Tr(\hat{a}_{k^{\prime}}\hat{a}_{k^{\prime}} \rho_{th}) 
=Tr(\hat{a}_{k^{\prime}}^{\dagger}\hat{a}_{k^{\prime}}^{\dagger} \rho_{th})=0 $ 
and $Tr(\hat{a}_{k^{\prime}}^{\dagger}\hat{a}_{k^{\prime}} \rho_{th})=\bar{n} 
$. Using Eq.~\eqref{eq:CM_gen} the 
covariance matrix of the two output modes $l$ and $m$ can be written as:
\begin{equation}\label{eq:term}
\sigma^{th}_{l,m} = \begin{pmatrix}
\sigma^{th}_\alpha& 0 & \sigma^{th}_\gamma & 0\\[4pt]
 0& \sigma^{th}_\alpha &0 & \sigma^{th}_\gamma \\[4pt]
\sigma^{th}_\gamma&0 & \sigma^{th}_\beta & 0\\[4pt]
0& \sigma^{th}_\gamma & 0 & \sigma^{th}_\beta
\end{pmatrix}
\end{equation}
with 
$\sigma^{th}_\alpha = |S_{l,k^{\prime}}|^2 \bar{n} + \frac{1}{2}$,
$\sigma^{th}_\beta = |S_{m,k^{\prime}}|^2 \bar{n} + \frac{1}{2}$, and
$\sigma^{th}_\gamma =  \bar{n} |S_{l,k^{\prime}}||S_{m,k^{\prime}}|$.

The values of $\gamma_x$ and $\gamma_p$ in this case are equal.
In Fig.~\ref{fig:regions} all possible thermal states lie on the line 
$\gamma_x=\gamma_p$ and thus no entanglement is possible according to the
criterion described in Eq.~\eqref{eq:determ}. In fact, in order for 
the modes to be entangled, $\gamma_x$ and $\gamma_p$ should at least have  
different signs \cite{Simon1999}. Although these states do not show 
entanglement, there are still quantum correlations between the output modes in 
the form of quantum discord.

\begin{figure}[t]
	\centering
	\includegraphics[scale=1.1]{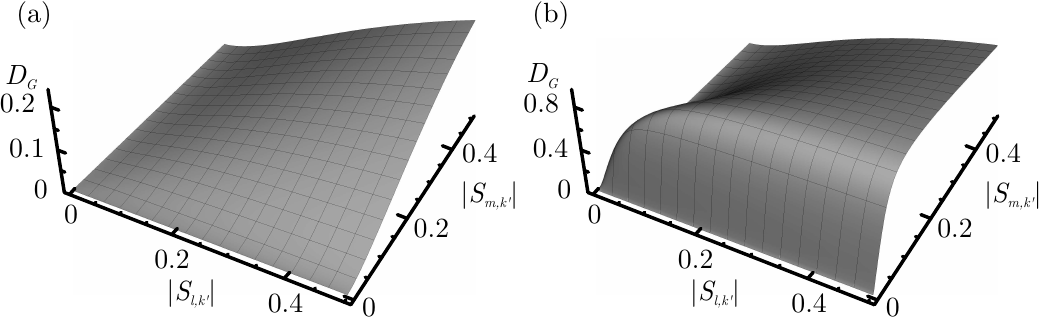}
	\caption{ \label{fig:discord} The dependence of the Gaussian discord, 
$D_G$, 
between modes $l$ and $m$ on the absolute values of the elements of the 
scattering matrix 
$|S_{l,k^{\prime}}|$ and $|S_{m,k^{\prime}}|$. The input mode, $k^\prime$, is 
in a thermal 
state with 
different average photon numbers (a)   $\bar{n}=1$, (b) $\bar{n}=10^3$. The 
measurement is performed on the mode~$l$.}

\end{figure}

We calculated the discord of the output covariance matrix for the case of input 
mode in the thermal state by substituting the output covariance matrix 
Eq.~\eqref{eq:term} into the formula for the Gaussian discord in 
Eq.~\eqref{eq:discord}. In Fig.~\ref{fig:discord} we plot the dependence of 
the Gaussian discord $D_G$ on 
the absolute values of the transmission coefficients from the mode $k^{\prime}$ 
to the modes $l$ and $m$. Notice that the discord $D$ is asymmetric 
against these coefficients since the measurement is performed on only one of 
the 
modes (on the mode $l$ in the Fig.~\ref{fig:discord}). The discord increases 
monotonously with $|S_{l,k^{\prime}}|$, 
but there can be a maximum in its dependence on $|S_{m,k^{\prime}}|$, the 
position of which is defined by the number of photons in the input mode 
(Fig.~\ref{fig:discord}b). At low photon numbers there is no maximum, and in 
that case the 
discord increases monotonously (Fig.~\ref{fig:discord}a). 
\begin{figure}[t]
	\centering
	\includegraphics[scale=1.1]{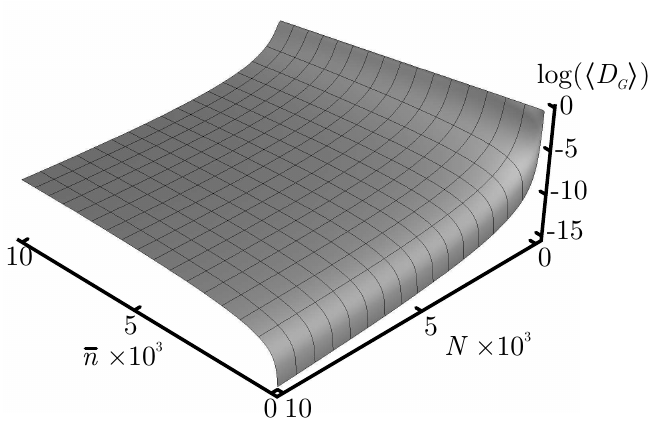}
	\caption{ \label{fig:Discord_n_N}Dependence of the average 
Gaussian discord, $\langle D_G \rangle$, 
on the number of output modes, $N$, and the number of photons, $\bar{n}$, in 
the 
thermal input state.}
\end{figure}

The mean discord obtained can be calculated noticing that, on average, the 
energy distributes equally among all the possible scattering channels and thus 
$\left \langle |S|^2 \right \rangle= 1/N$. If we assume that the scattered 
light follows a Rayleigh distribution we have $\left \langle |S|^2 \right 
\rangle = 2 \left \langle |S| \right \rangle$ \cite{Akkermans}, which can be 
substituted into Eq.~\eqref{eq:discord} to get the average amount of discord in 
a pair of output modes in this configuration. As shown in 
Fig.~\ref{fig:Discord_n_N}, $\left \langle D_G \right \rangle$ increases 
monotonically with $\bar{n}$, but decreases monotonically with $N$. As a 
consequence 
the best conditions to observe the discord generated by multiple scattering of 
a thermal state of light are obtained for an intense light signal scattering 
over a system with a small number of channels. Therefore we suggest that 
light scattering from systems showing Anderson localization \cite{Sheng, 
Akkermans} will show a significant amount of quantum discord.

\section{Conclusions}

The search for a universal criterion that captures all the 
nuances of non-classical correlations is an object of ongoing 
intensive discussion \cite{Ferraro, Saleh}. The presented results contribute 
to this debate by providing an illustration of the differences between 
various measures of quantum correlations, such as the correlation function 
based on intensity fluctuations $\mathcal{C}$, entanglement and discord. 
We calculated the covariance matrix of two arbitrary output modes of the light 
elastically scattered  by a disordered material for different states of the 
input mode, and analysed their correlation properties. Surprisingly, the 
results show that if the input is a thermal state then any two output modes 
will 
be (Gaussian) discorded, a signature of the quantum character of light. 
Moreover, it turns out that coherent states are the \emph{only} Gaussian input 
that do not produce quantum correlations, as measured by any of the quantities 
considered here.

It is known that the propagation of light through a scattering medium is 
modified by quantum interference when the input state is entangled 
\cite{Silberberg}, but the effects of quantum discord on light propagation are 
still a largely unexplored subject. Quantum discord appears naturally from 
the multiple scattering of thermal light, even for large photon numbers. Such 
macroscopic effects can potentially be exploited to develop novel imaging 
techniques. Since the amount of expected quantum discord grows 
when the number of scattering channels is small, these effects will play a role 
especially in the case of strongly scattering materials, where the 
dimensionless 
conductance $g$ is small~\cite{Akkermans}. In particular we expect it to have 
an 
effect for systems that show Anderson localization.

Finally, when light undergoes multiple scattering, even very weak 
nonlinearities 
can have a dramatic effect~\cite{Wellens}. Thus diffusion through a nonlinear 
system will produce a very rich landscape of possible output states, opening 
the 
possibility to generate multimode entanglement from classical input light.

\section{Acknowledgements}
We are grateful to M. Paternostro, D. Browne, and M. Williamson for insightful 
discussions. 
JA acknowledges support by EPSRC (EP/M009165/1). JB acknowledges support from 
the Leverhulme Trust's Philip Leverhulme Prize. IS acknowledges support from 
EPSRC through the Centre of Doctoral Training in Metamaterials ($\text{XM}^2$).

\end{document}